\documentclass[prx,aps,twocolumn,superscriptaddress]{revtex4-2}

\usepackage[pdftex]{graphicx}
\usepackage{amsbsy,amssymb,amsmath,bm,mathtools}
\usepackage{amsfonts}
\usepackage{mathrsfs}
\usepackage{makecell}
\usepackage{physics}

\usepackage{xspace}
\usepackage{bm}
\usepackage{color}
\usepackage{url}
\usepackage[section]{placeins}
\usepackage[colorlinks=true,linkcolor=blue,citecolor=blue]{hyperref}

\usepackage{tikz}

\begin{document} 

\title{Emergent Loop Geometry in Gauge-Equivariant Graph Neural Networks}

\author{Ali Rayat}
\affiliation{Department of Physics, University of Virginia, Charlottesville, VA 22904, USA}

\author{Gia-Wei Chern}
\affiliation{Department of Physics, University of Virginia, Charlottesville, VA 22904, USA}

\date{\today}

\begin{abstract}
Gauge-equivariant graph neural networks generate ordered products of gauge links through local message passing, but the geometry of these internal structures is not captured by the receptive field alone. We investigate the statistical ensemble of closed lattice walks generated by recursive node- and edge-feature operations, independently of network training. Analytical and numerical results reveal a separation between rapid growth of contour length and much slower spatial expansion: deeper compositions generate increasingly long paths through repeated traversal of relatively compact regions. Loop expansions of fermionic observables provide a physical interpretation of this hierarchy, showing how spatial locality remains compatible with high-order dependence on nearby gauge fields. These findings identify complementary roles of network depth in extending spatial reach and enriching the representations constructed within that reach, providing a geometrical perspective on learning lattice gauge observables.
\end{abstract}

\maketitle

\section{introduction}

\label{sec:intro}

Non-Abelian gauge theories provide a unifying framework for systems governed by local symmetry constraints. In high-energy physics, they underpin the Standard Model and, in particular, quantum chromodynamics, where strongly coupled gauge fields give rise to confinement and other intrinsically nonperturbative phenomena~\cite{wilson1974,kogut1975,creutz1983,rothe2012}. Gauge structures also emerge in condensed-matter systems, particularly in fractionalized and strongly constrained phases such as quantum spin liquids~\cite{wen2004,lee2006,savary2017,zhou2017}. Quantum link models and related finite-dimensional formulations connect lattice gauge theories to atomic, molecular, and optical systems, where gauge constraints and gauge--matter dynamics can be engineered in quantum simulators~\cite{chandrasekharan1997,wiese2013,zohar2016,tagliacozzo2013,zhou2022}. These diverse settings motivate computational approaches that efficiently represent gauge correlations while preserving local symmetry.

Machine learning has emerged as a promising tool for addressing computational problems in lattice gauge theories~\cite{cranmer2023,tomiya2025}. Applications span accelerated Monte Carlo sampling and effective-action learning, neural representations of quantum states, lattice-QCD preconditioning, and prediction of gauge-invariant and gauge-covariant quantities~\cite{albergo2019,kanwar2020,boyda2021,abbott2022,luo2021,apte2024,Spriggs2026,lehner2023,lehner2024,favoni2022,nagai2025}. A central development across these applications is the explicit incorporation of gauge symmetry into the learned representations. Gauge invariance can be enforced by working with invariant variables, such as Wilson loops, or by imposing local gauge constraints directly in neural quantum-state constructions~\cite{nagai2023,luo2023,rayat2026-a}. Gauge-equivariant convolutional networks, normalizing flows, and related constructions retain features with prescribed local transformation laws, using symmetry-compatible operations and, where appropriate, parallel transport to combine information from different sites. These approaches offer complementary ways to preserve gauge structure while learning the correlations relevant to physical observables.

These developments are part of a broader effort to incorporate physical structure and symmetry into neural networks. Graph neural networks (GNNs), particularly message-passing architectures, iteratively propagate and aggregate information carried by nodes and edges over local neighborhoods~\cite{scarselli2009,gilmer2017,hamilton2017,xu2019}. Equivariant neural networks (ENNs) enforce prescribed transformation laws and have become important tools in molecular and materials modeling~\cite{cohen2016,cohen2017,cohen2018,weiler2018,schutt17,schutt2018,batatia2022,Geiger22}. Combining equivariance with message passing allows such symmetry constraints to be maintained as information is gathered and composed across a graph.

In many of these applications, equivariance concerns \emph{global} symmetries, for which a common transformation acts on the entire system. Gauge symmetry instead permits independent transformations at every lattice site, requiring information to be transported consistently between local gauge frames. In our previous work, we introduced a general formulation of non-Abelian gauge equivariance within a graph message-passing framework~\cite{rayat2026}. Matrix-valued node and edge features transform covariantly under local gauge transformations, and their update rules combine local feature products with gauge-covariant transport. The architecture supports learning gauge-invariant and gauge-covariant quantities in pure gauge, gauge--matter, and dynamical gauge systems.

A distinctive feature of this construction is that the gauge-link products generated by message passing admit a geometrical interpretation as paths and loops on the lattice~\cite{rayat2026}. Successive updates concatenate elementary transports and compose closed-loop structures analogous to Wilson loops, without prescribing their shapes or sizes in advance~\cite{wilson1974,favoni2022}. These paths can retrace links, revisit sites, and contain nested excursions, so their contour lengths need not track their spatial extent. The conventional graph-theoretic receptive field therefore captures only part of the structure developed through depth: it bounds spatial reach but does not describe the complexity of the paths composed within that region. Understanding this distinction requires a statistical characterization of the generated loop ensemble.

In this work, we study the combinatorial and geometrical structures produced by the propagation rules, independently of network training. We characterize loop-length distributions and their dependence on recursion depth and spatial dimension, together with measures of spatial extent, shape, site exploration, and length diversity. The central result is a separation between exponential growth of the mean contour length and much slower growth of spatial extent over the depths studied. Correlations among these quantities show how increasingly long paths accumulate traversals within relatively compact regions while retaining substantial geometrical diversity. Through explicit loop expansions of fermionic observables, we relate this hierarchy to the distinction between hopping order and physical locality, and discuss its implications for representing complex gauge-field dependence through local message passing.

The remainder of this paper is organized as follows. We begin by reviewing the gauge-equivariant message-passing rules and their path construction in Sec.~\ref{sec:gnn}, then characterize the resulting loop-length statistics and spatial geometry in Sec.~\ref{sec:loop_geometry}. Their physical interpretation through loop expansions and locality is developed in Sec.~\ref{sec:implications}. We conclude in Sec.~\ref{sec:conclusion} with a summary and implications for GNN design.
Appendix~\ref{app:loop_generation} details the recursive sampling procedure and generating-function analysis, while Appendix~\ref{app:loop_correlations} examines correlations between contour length and individual-loop geometry.

\section{Gauge-equivariant message passing}
\label{sec:gnn}

\begin{figure*}[t]
\centering
\includegraphics[width=1.99\columnwidth]{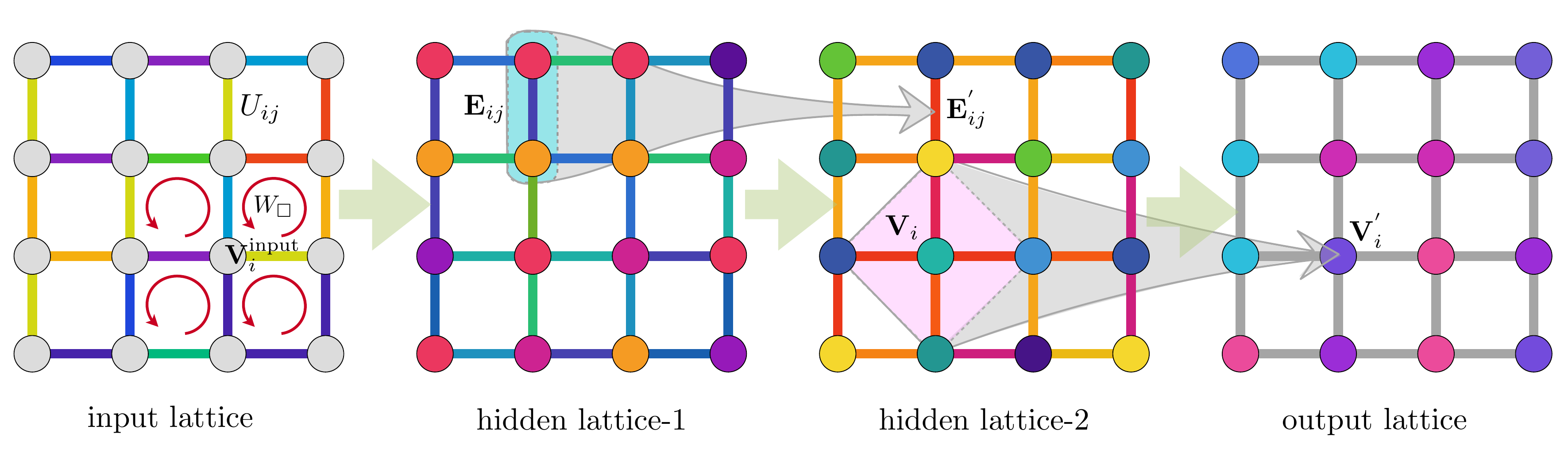}
\caption{Schematic of the graph neural network (GNN) and message-passing architecture. The input lattice is represented as a graph with information carried by both nodes and edges. Starting from the input edge variables $U_{ij}$ and local node features $\mathbf{V}_i^{\rm input}$, successive message-passing layers generate hidden node features $\mathbf{V}_i$ and edge features $\mathbf{E}_{ij}$. At each layer, an edge feature $\mathbf{E}_{ij}$ is updated using information from the edge itself and the two nodes at its endpoints (top), while a node feature $\mathbf{V}_i$ is updated by combining its own features with information aggregated from its four nearest-neighbor nodes (bottom). Repeated message passing therefore propagates information over increasingly extended regions of the lattice, producing the final node and edge representations at the output layer. Different colors schematically distinguish feature channels and are not intended to represent their gauge-transformation properties.}
    \label{fig:gnn}
\end{figure*}

The loop structures analyzed in this work arise from the recursive propagation rules of the gauge-equivariant graph neural network introduced in Ref.~\cite{rayat2026}. We review the architecture here from a geometrical perspective, following how local message-passing operations compose elementary gauge links into increasingly elaborate paths. Figure~\ref{fig:gnn} summarizes the construction. The transformation laws specify which features can be combined, while the allowed products determine the path structures generated from one layer to the next.

Consider a lattice gauge configuration specified by group-valued variables $U_{ij}\in G$ on oriented nearest-neighbor links $(ij)$. The link variable $U_{ij}$ represents parallel transport from site $j$ to site $i$ and transforms under independent local gauge transformations $g_i\in G$ as~\cite{rothe2012}
\begin{equation}
U_{ij}\rightarrow g_i U_{ij}g_j^\dagger .
\label{eq:gauge_link}
\end{equation}
The lattice provides a natural graph representation, with sites as nodes and gauge links as edges. Message passing preserves this underlying graph while building collections of embedded features on its nodes and edges. Although the connectivity remains local, the features can encode products involving paths that extend beyond a single bond or plaquette.

At layer $n$, we associate matrix-valued feature collections $\mathbf{V}_i^{(n)}=\{V_{i,\alpha}^{(n)}\}$ with each node $i$ and $\mathbf{E}_{ij}^{(n)}=\{E_{ij,\beta}^{(n)}\}$ with each oriented edge $(ij)$. The channel indices $\alpha$ and $\beta$ distinguish features within these collections. Their transformation laws mirror those of closed and open gauge transports,
\begin{equation}
\mathbf{V}_i^{(n)}
\rightarrow
g_i\mathbf{V}_i^{(n)}g_i^\dagger,
\qquad
\mathbf{E}_{ij}^{(n)}
\rightarrow
g_i\mathbf{E}_{ij}^{(n)}g_j^\dagger .
\label{eq:feature_transform}
\end{equation}
A node feature is thus anchored in the gauge frame of a single site, whereas an edge feature connects the frames at its two endpoints. Reversing the edge orientation corresponds to Hermitian conjugation, $\mathbf{E}_{ji}^{(n)}=\mathbf{E}_{ij}^{(n)\dagger}$. These endpoint properties are preserved throughout the network, even as the products represented by individual features become longer and more complex.

The input layer, $n=0$, supplies the elementary building blocks. Each edge contains the corresponding gauge link,
\begin{equation}
\mathbf{E}_{ij}^{(0)}=\{U_{ij}\},
\label{eq:edge_initialization}
\end{equation}
while node features are initialized from plaquette holonomies based at that site. For a plaquette with ordered vertices $(i,j,k,l)$, one may take $V_{i,\alpha}^{(0)}=U_{ij}U_{jk}U_{kl}U_{li}$, with neighboring plaquettes and their orientations furnishing different channels $\alpha$. Here and below, geometrical paths are listed in the left-to-right order of link indices; the corresponding matrices act on color vectors from right to left, transporting them in the reverse direction. The input therefore consists of open paths of one lattice bond and elementary closed plaquette loops. The plaquette products remain untraced: retaining their matrix structure allows them to be transported and multiplied in subsequent layers before gauge-invariant quantities are formed.

Each message-passing layer first gathers information through symmetry-compatible operations and then combines the resulting messages into the next set of features. We separate these two steps to expose the operations responsible for generating paths. The three node-message contributions are denoted by $\mathbf{X}_{i,r}^{(n)}$, with $r=1,2,3$. Here and below, $\leftarrow$ specifies the operation that generates a collection of message features, and products of boldface arrays denote the corresponding products among their constituent matrices.

The first two node messages use only features already anchored at $i$,
\begin{align}
\mathbf{X}_{i,1}^{(n)}
&\leftarrow
\mathbf{V}_i^{(n)},
\label{eq:node_message_1}\\
\mathbf{X}_{i,2}^{(n)}
&\leftarrow
\mathbf{V}_i^{(n)}\mathbf{V}_i^{(n)}.
\label{eq:node_message_2}
\end{align}
The first carries existing features forward, preserving access to structures generated at earlier layers. The second forms pairwise products, allowing those structures to be composed locally. Both preserve the node transformation law: in a product of two node features, the adjacent factors $g_i^\dagger g_i$ cancel, leaving the same conjugation by $g_i$ as for either factor separately.

The third message incorporates neighboring nodes. Let $\mathcal N(i)$ denote the nearest neighbors of $i$, comprising the four adjacent sites on the square lattice in Fig.~\ref{fig:gnn}. A feature $\mathbf{V}_j^{(n)}$ transforms in the gauge frame at $j$, so it must be brought into the frame at $i$ before it can be combined with the local features. The connecting edge features provide this transport,
\begin{equation}
\mathbf{X}_{i,3}^{(n)}
\leftarrow
\sum_{j\in\mathcal{N}(i)}
\mathbf{E}_{ij}^{(n)}
\mathbf{V}_j^{(n)}
\mathbf{E}_{ji}^{(n)} .
\label{eq:node_message_3}
\end{equation}
Each contribution obeys $\mathbf{E}_{ij}^{(n)}\mathbf{V}_j^{(n)}\mathbf{E}_{ji}^{(n)}\rightarrow g_i\bigl[\mathbf{E}_{ij}^{(n)}\mathbf{V}_j^{(n)}\mathbf{E}_{ji}^{(n)}\bigr]g_i^\dagger$: the transformations associated with $j$ cancel inside the product. The transported messages can therefore be aggregated with the two local contributions without violating gauge covariance.

The gathered messages enter a gauge-compatible node update,
\begin{equation}
\mathbf{V}_i^{(n+1)}
=
\mathcal{U}_V
\biggl[
\sum_{r=1}^{3}\mathbf{X}_{i,r}^{(n)}
\biggr].
\label{eq:node_update}
\end{equation}
Here $\mathcal{U}_V$ represents the channel mixing, normalization, and nonlinear processing used in the full GNN. These operations determine how the available features are weighted and embedded. For the geometrical analysis, we track the path connectivity supplied by the message products in Eqs.~\eqref{eq:node_message_1}--\eqref{eq:node_message_3}, separately from this subsequent numerical processing.

Edge updates complement the node construction by allowing the transporters themselves to acquire additional structure. An oriented edge $(ij)$ gathers information from its current features and from the nodes at both endpoints. The three edge-message types are
\begin{align}
\mathbf{Y}_{ij,1}^{(n)}
&\leftarrow
\mathbf{E}_{ij}^{(n)},
\label{eq:edge_message_1}\\
\mathbf{Y}_{ij,2}^{(n)}
&\leftarrow
\mathbf{V}_i^{(n)}\mathbf{E}_{ij}^{(n)}
+
\mathbf{E}_{ij}^{(n)}\mathbf{V}_j^{(n)},
\label{eq:edge_message_2}\\
\mathbf{Y}_{ij,3}^{(n)}
&\leftarrow
\mathbf{V}_i^{(n)}
\mathbf{E}_{ij}^{(n)}
\mathbf{V}_j^{(n)}.
\label{eq:edge_message_3}
\end{align}
The first carries the existing edge representation forward. The two products in the second dress the edge at either endpoint, while the third attaches node features at both endpoints. Their ordering ensures that all internal gauge transformations cancel, leaving $\mathbf{Y}_{ij,r}^{(n)}\rightarrow g_i\mathbf{Y}_{ij,r}^{(n)}g_j^\dagger$ for each $r$. Thus, dressing an edge changes the information it carries while preserving its endpoint transformation law.

The edge features for the next layer are then obtained through
\begin{equation}
\mathbf{E}_{ij}^{(n+1)}
=
\mathcal{U}_E
\biggl[
\sum_{r=1}^{3}\mathbf{Y}_{ij,r}^{(n)}
\biggr].
\label{eq:edge_update}
\end{equation}
As in the node update, $\mathcal U_E$ combines and processes the gathered features. The geometrical role of the edge messages is already apparent in their matrix products: a feature assigned to a nearest-neighbor edge can contain excursions through neighboring loops while retaining the same two endpoints.

The complete update cycle can now be read as a coupled construction of closed and open walks. Node features are associated with closed walks based at a site, and edge features with open walks connecting neighboring sites. Whenever endpoints match, matrix multiplication concatenates the corresponding paths. In particular, $\mathbf{V}_i^{(n)}\mathbf{V}_i^{(n)}$ joins two loops at a common base point, while $\mathbf{E}_{ij}^{(n)}\mathbf{V}_j^{(n)}\mathbf{E}_{ji}^{(n)}$ connects a loop at $j$ to the reference site $i$ through an outgoing path and its reverse. The edge product $\mathbf{V}_i^{(n)}\mathbf{E}_{ij}^{(n)}\mathbf{V}_j^{(n)}$ similarly attaches loops to both ends of an open path. Closed walks therefore dress the open paths through edge updates, and these dressed paths enter the construction of new closed walks through subsequent node updates.

This mutual recursion is the source of the loop ensemble studied below. Starting from elementary links and plaquettes, successive layers combine previously constructed paths into walks with a broad range of contour lengths and spatial geometries. Concatenation can add many traversals without reaching new sites, while transport introduces excursions that may revisit regions already explored. As a result, contour length and spatial extent need not grow together, even though both emerge from the same local update rules. We first illustrate these mechanisms through representative short loops whose construction can be followed explicitly, before turning to the statistical properties of the much larger ensemble generated at greater depth.

\begin{figure}[t]
\centering
\includegraphics[width=0.99\columnwidth]{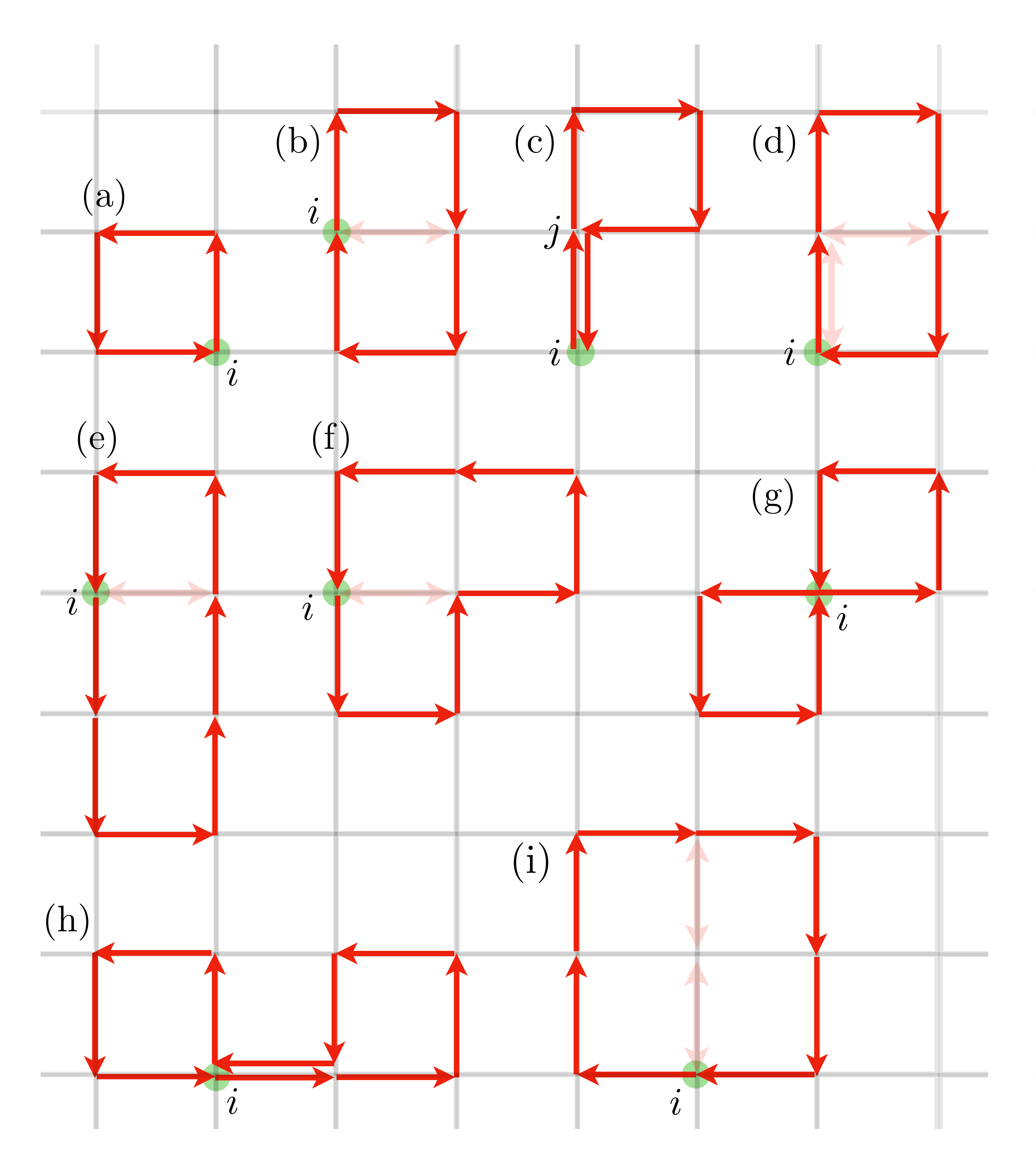}
\caption{Examples of closed lattice paths generated by gauge-equivariant message passing and represented by node features $\mathbf{V}_i$. The green circle marks the reference node $i$ at which each loop is anchored, while the red arrows indicate the ordered sequence of lattice links forming the path. Starting from elementary plaquette loops, successive node and edge message-passing operations generate increasingly extended and geometrically complex closed paths, including paths that retrace links or revisit lattice sites. Panels (a)--(i) show representative short loops generated at different stages of this recursive construction. The corresponding message-gathering operations for these examples are discussed in the main text.}
    \label{fig:short-loop}
\end{figure}

\begin{figure*}[t]
\centering
\includegraphics[width=1.99\columnwidth]{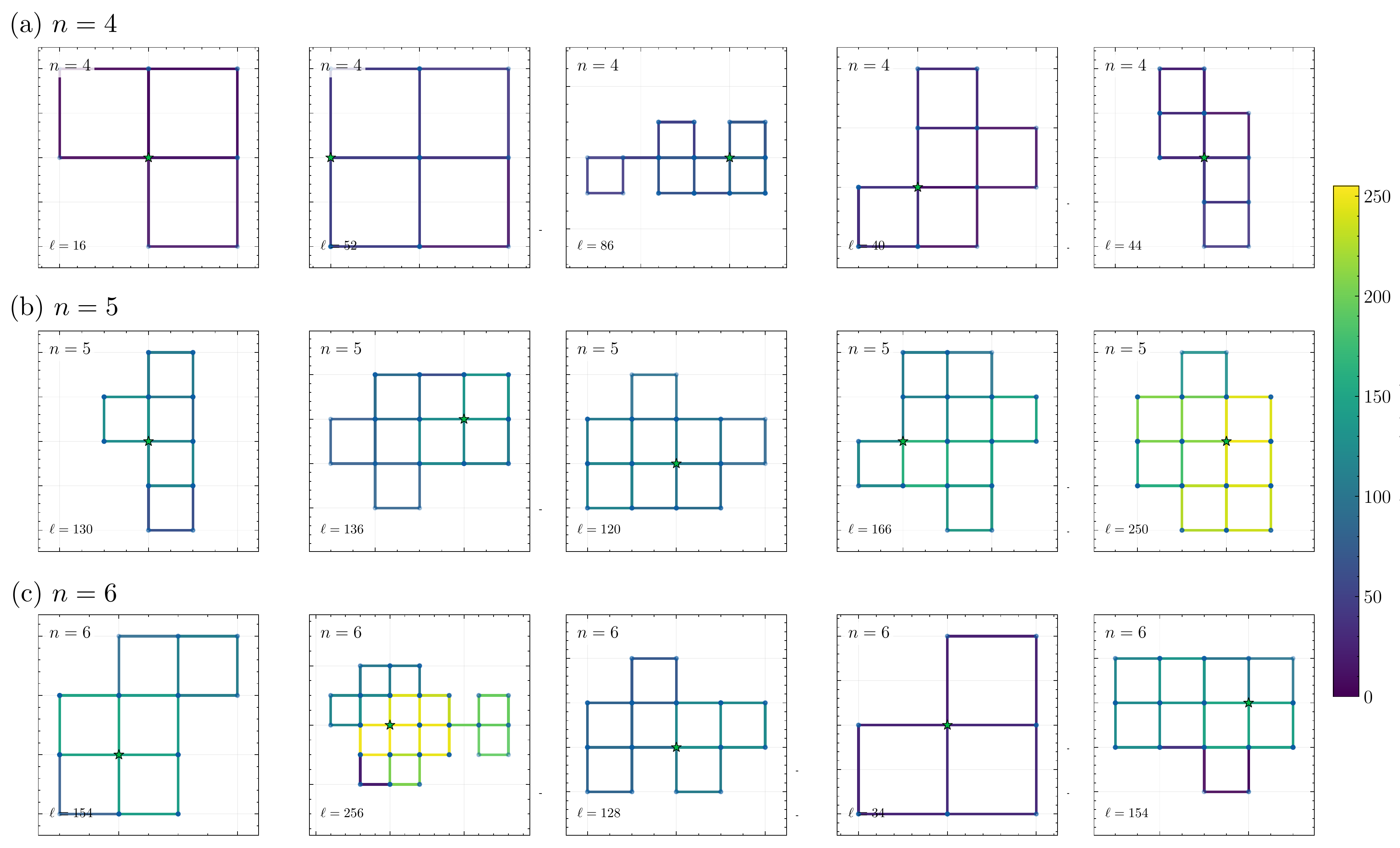}
\caption{Representative closed paths generated by the recursive loop-sampling algorithm. Rows (a)--(c) show realizations at recursion depths $n=4$, $5$, and $6$, respectively; $\ell$ denotes the full contour length of each path. The green star marks the reference site where the path begins and ends. Color indicates the traversal step $k$, progressing from the beginning to the end of each path. Unlike the short examples in Fig.~\ref{fig:short-loop}, these realizations are generated by stochastic selection of node and edge message operations, without prescribing their shapes. Paths generated at the same depth vary substantially in length and geometry and can traverse links repeatedly, revisit sites, and contain nested or overlapping subloops.}
\label{fig:long-loops}
\end{figure*}

Figure~\ref{fig:short-loop} illustrates how the message-passing rules generate several representative closed paths. The green circle marks the reference site $i$ associated with the node feature, and the arrows indicate the ordered traversal of lattice links. Even these short examples reveal that a generated path need not be a simple geometrical loop: elementary loops can be concatenated, bonds can be traversed repeatedly, and a walk can return to the same site at different stages. Its spatial footprint alone therefore does not specify either the full sequence of traversals or the message operations from which it arose.

The elementary square in Fig.~\ref{fig:short-loop}(a) is present at the input layer as a plaquette feature $V_{i,\alpha}^{(0)}$. The first extensions arise from the local product in message type $\mathbf{X}_{i,2}$. For example, multiplying two adjacent plaquette features with a common reference site, $V_{i,1}^{(0)}V_{i,2}^{(0)}$, generates the structure in panel (b) at $n=1$. With the indicated orientations and ordering, the opposite traversals of the shared bond occur consecutively and cancel, $U_{ij}U_{ji}=I$, leaving the length-six perimeter of the two plaquettes. Such cancellations simplify the schematic paths in Fig.~\ref{fig:short-loop}; the statistical analysis below instead counts every traversal in the original generated path.

Panel (c) shows how the nearest-neighbor message $\mathbf{X}_{i,3}$ produces a different kind of extension. A plaquette based at a neighboring site $j$ is transported to $i$ through the product $E_{ij}^{(0)}V_{j,\alpha}^{(0)}E_{ji}^{(0)}$. The outgoing and returning traversals of the $i$--$j$ bond are separated by the plaquette factor $V_{j,\alpha}^{(0)}$, so they do not cancel within this untraced matrix product. The path follows the connecting bond, completes the neighboring plaquette, and returns along the same bond. Its contour length thus includes a retraced segment that adds no new area to the enclosed plaquette.

The new structure in panel (d) follows by multiplying this transported loop by an elementary plaquette based at $i$, through another type-$\mathbf{X}_{i,2}$ operation. Schematically, the product is $V_{i,\mathrm{(c)}}^{(1)}V_{i,\alpha}^{(0)}$, with the elementary plaquette retained at layer $n=1$ through the carry-forward message and oriented to occupy the lower square in panel (d). At the junction between the factors, the return segment of the transported loop cancels against the initial segment of the elementary plaquette. This exposes another oppositely oriented pair along the shared horizontal bond, which also cancels. The resulting path follows the length-six rectangular perimeter, now based at its lower-left corner.

Panels (b) and (d) thus illustrate an important distinction between a traced Wilson loop and its untraced holonomy. When placed on the same oriented rectangular contour, they differ only in the choice of base point. Their holonomies are related by conjugation with the parallel transporter between the two base points and consequently have exactly the same trace. The untraced matrices, however, are generally different and transform in the gauge frames of their respective reference sites. Equality of the standalone traces therefore does not make these features interchangeable in subsequent message passing: their composition with other node features depends on the full matrices and their common anchoring. Retaining matrix-valued holonomies preserves this information for further path construction.

\begin{figure*}[t]
\centering
\includegraphics[width=1.99\columnwidth]{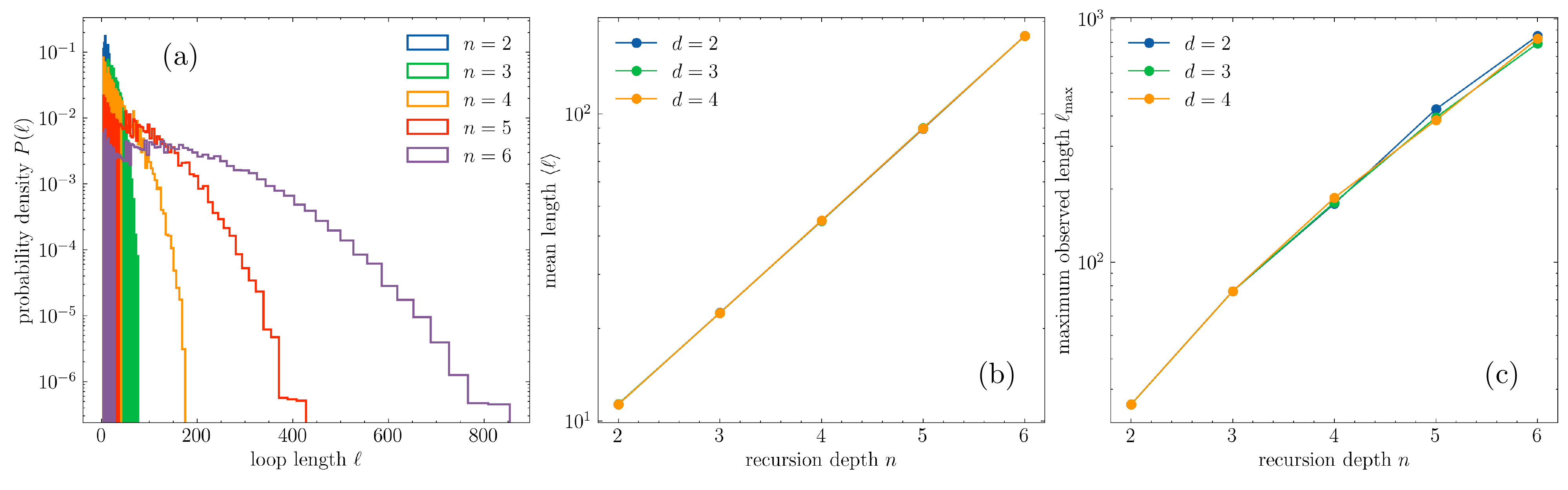}
\caption{Loop-length statistics of recursively generated closed paths.
(a) Semilog plot of the loop-length distribution $P_n(\ell)$ in $d=2$ for recursion depths $n=2,\ldots,6$. The tails bend downward, indicating that their decay steepens with $\ell$ and is faster than a single exponential over the displayed range.
(b) Mean loop length $\langle\ell\rangle$ and
(c) maximum observed loop length $\ell_{\max}$ as functions of recursion depth for $d=2,3,4$. Both increase rapidly with depth; the maximum is the largest length in the sampled loops, rather than the maximum allowed by the recursion.}
\label{fig:loop-length-hist}
\end{figure*}

Further compositions build on these structures, using the appropriate placements and orientations of the constituent loops. The three-square path in panel (e) can be obtained by combining features of the types shown in panels (b) and (c) through another type-$\mathbf{X}_{i,2}$ operation, schematically $V_{i,\mathrm{(b)}}^{(1)}V_{i,\mathrm{(c)}}^{(1)}$. Panel (f) similarly joins the loop in panel (b) to an elementary plaquette based at the same site, $V_{i,\mathrm{(b)}}^{(1)}V_{i,\alpha}^{(0)}$. The diagonal arrangement in panel (g) illustrates another possibility: local multiplication requires a common base point, but does not require the constituent plaquettes to share a bond or lie along the same axis. Panel (h) combines the transported loop of panel (c) with an elementary plaquette through the same local product operation, but with a different relative placement from panel (d), so the connecting excursion remains. Together, these examples show how the geometry and ordering of the constituent paths determine whether concatenation retains retraced segments or allows them to cancel.

Finally, the $2\times2$ square in panel (i) follows directly from a type-$\mathbf{X}_{i,2}$ product of two rectangular features of the type constructed in panel (d). The two rectangles occupy the left and right halves of the square and share the reference site $i$ at the midpoint of its lower edge. With the appropriate ordering, their opposite traversals of the shared two-bond segment become adjacent in the product and cancel successively. The remaining path follows the length-eight outer perimeter. This construction makes explicit how repeated local multiplication can assemble a larger Wilson loop from previously generated features, while the same update rule also produces the retraced and multiply connected walks illustrated in the other panels.

These examples follow selected compositions whose effects can be traced by hand. To characterize the much larger ensemble available at greater depth, we generate paths by sampling the recursive choices. Starting from a node feature at depth $n$, we choose uniformly among the three node-message operations. For an edge feature, we choose uniformly among the four products represented by its three messages, counting the two terms of $\mathbf{Y}_2$ separately. Every factor in the chosen product is then generated recursively until the construction reaches the input plaquettes and links. In a transported node message, the outgoing path is sampled once and its exact reverse supplies the return path.

The resulting construction has the structure of an AND/OR derivation tree: choosing a message operation selects one alternative, while forming its product requires all constituent factors~\cite{ChangSlagle1971}. Each realization therefore selects a derivation subtree specifying one closed path, as illustrated in Fig.~\ref{fig:tree-schematic}. Appendix~\ref{app:loop_generation} gives the full sampling procedure. This construction follows the available message products without prescribing the final path shape or using learned network weights.

Figure~\ref{fig:long-loops} shows representative realizations at depths $n=4$, $5$, and $6$. Paths generated at the same depth can differ substantially in contour length because carry-forward operations preserve existing structures, whereas product operations combine several recursively generated factors. Despite their length, many realizations remain concentrated within a relatively small region, accumulating steps through repeated visits, retraced bonds, and nested or overlapping excursions. All link traversals are retained in these numerical paths, including the immediate backtracking pairs removed from some of the short-loop schematics. The resulting variation in both length and spatial organization motivates the statistical characterization that follows.

\section{Statistical geometry of generated loops}
\label{sec:loop_geometry}

We characterize the generated loops in dimensions $d=2,3,4$ at recursion depths $n=2,\ldots,6$. The contour length $\ell$ counts all elementary link traversals, including repeated and immediately retraced links. The statistics refer to the random recursion specified above; uniform selection of local message products does not imply uniform sampling of distinct final loops.

Figure~\ref{fig:loop-length-hist}(a) shows that the length distribution $P_n(\ell)$ broadens and shifts toward larger $\ell$ with increasing depth. The growth of its mean follows exactly from the additive lengths of the message products. Let $\bar\ell_V(n)$ and $\bar\ell_E(n)$ denote the mean node and edge lengths after $n$ updates. The three node choices contribute lengths $\ell_V$, $2\ell_V$, and $\ell_V+2\ell_E$, while the four edge products contribute $\ell_E$, $\ell_E+\ell_V$, $\ell_E+\ell_V$, and $\ell_E+2\ell_V$. Averaging these choices gives
\begin{align}
\bar\ell_V(n) &= \frac{4}{3}\bar\ell_V(n-1)+\frac{2}{3}\bar\ell_E(n-1),\\
\bar\ell_E(n) &= \bar\ell_V(n-1)+\bar\ell_E(n-1).
\label{eq:loop_mean_recursion}
\end{align}
These first-moment equations are exact under the stated sampling rules, including the equality of the outgoing and return path lengths in the transported node message. With $\bar\ell_V(0)=4$ and $\bar\ell_E(0)=1$, their solution is
\begin{equation}
\langle\ell\rangle=\bar\ell_V(n)=\frac{14}{5}\,2^n+\frac{6}{5}\,3^{-n}.
\label{eq:loop_mean_solution}
\end{equation}
The asymptotic growth $\langle\ell\rangle\sim(14/5)2^n$ agrees with Fig.~\ref{fig:loop-length-hist}(b). Neither the input lengths nor the replacement probabilities depend on $d$, so the full length distribution is dimension independent under this recursion, consistent with the nearly coincident numerical results. Appendix~\ref{app:loop_generation} gives the corresponding probability-generating functions.

\begin{figure*}[t]
\centering
\includegraphics[width=1.99\columnwidth]{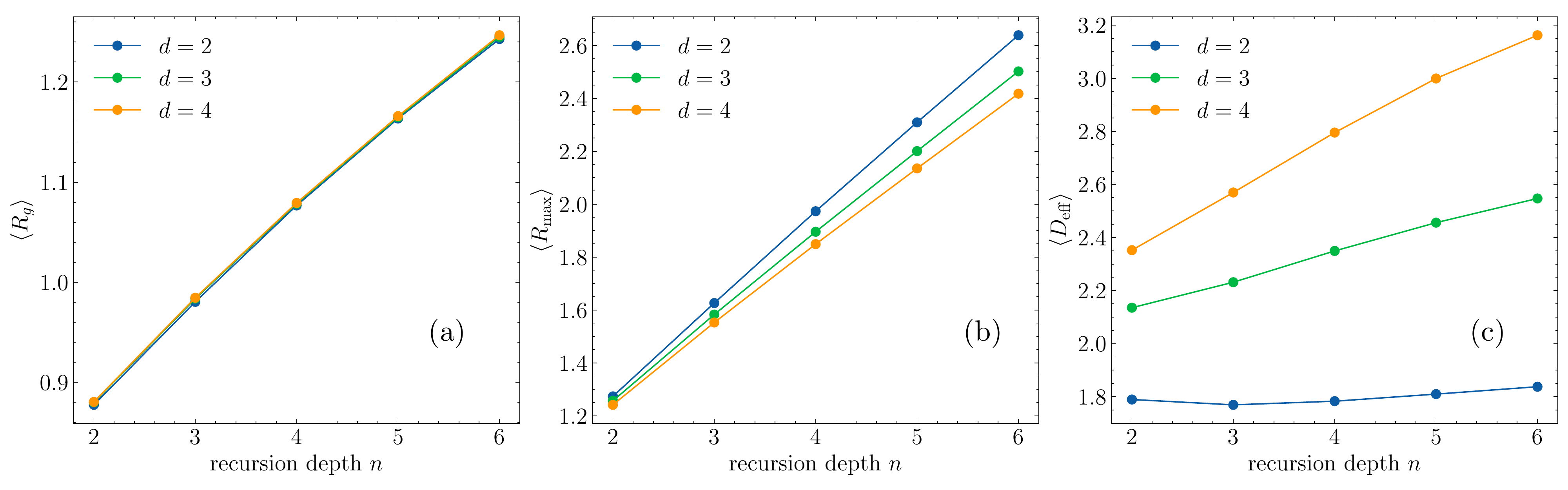}
\caption{Geometric characterization of generated loops in dimensions $d=2$, $3$, and $4$ as a function of recursion depth $n$. Shown are the ensemble-averaged (a) radius of gyration $\langle R_g\rangle$, (b) maximum distance from the loop center of mass $\langle R_{\max}\rangle$, and (c) effective dimension $\langle D_{\mathrm{eff}}\rangle$. The radius of gyration is nearly independent of $d$, while the maximum excursion shows a modest dimensional dependence. Both spatial measures grow much more slowly than the mean contour length in Fig.~\ref{fig:loop-length-hist}(b). The effective dimension remains nearly constant for $d=2$ but increases with depth for $d=3$ and $4$.}
\label{fig:loop-geometry}
\end{figure*}

The maximum permitted length follows by selecting the longest message products. Denoting the maximum node and edge lengths by $\ell_{V,\max}(n)$ and $\ell_{E,\max}(n)$, respectively, we obtain
\begin{align}
\ell_{V,\max}(n) &= \max\bigl\{2\ell_{V,\max}(n-1),\nonumber\\
&\qquad \ell_{V,\max}(n-1)+2\ell_{E,\max}(n-1)\bigr\},\\
\ell_{E,\max}(n) &= 2\ell_{V,\max}(n-1)+\ell_{E,\max}(n-1),
\label{eq:loop_max_recursion}
\end{align}
with $\ell_{V,\max}(0)=4$ and $\ell_{E,\max}(0)=1$. From $n=2$ onward, the second choice sets the node maximum, giving
\begin{equation}
\ell_{V,\max}(n)=\frac{17}{6}\,3^n+\frac{1}{2}(-1)^n,\qquad n\geq 1.
\label{eq:loop_max_solution}
\end{equation}
The permitted maximum therefore grows as $3^n$, faster than the mean. Figure~\ref{fig:loop-length-hist}(c) shows the maximum observed in a finite sample, which can lie well below this bound because the maximizing choices are rarely realized. Its value depends on the sample size.

The exponential growth of the mean with depth does not imply an exponential distribution in length. The tails in Fig.~\ref{fig:loop-length-hist}(a) bend downward on semilog axes, indicating that their decay steepens over the displayed range. Since each finite-depth distribution has bounded support, this observation does not establish an asymptotic tail law. The generating-function recursion in Eq.~\eqref{eq:loop_length_pgf} provides a sampling-free means of evaluating $P_n(\ell)$.

To compare contour growth with spatial expansion, let $\mathbf r_k$ be the unwrapped lattice position after step $k$ and define $\mathbf r_{\mathrm{cm}}=\ell^{-1}\sum_{k=0}^{\ell-1}\mathbf r_k$. The repeated final endpoint is omitted. The radius of gyration and maximum excursion are~\cite{rubinstein2003,rudnick2004}
\begin{equation}
\begin{aligned}
R_g^2 &= \frac{1}{\ell}\sum_{k=0}^{\ell-1}
|\mathbf r_k-\mathbf r_{\mathrm{cm}}|^2,\\
R_{\max} &= \max_{0\leq k<\ell}
|\mathbf r_k-\mathbf r_{\mathrm{cm}}|.
\end{aligned}
\label{eq:loop_spatial_size}
\end{equation}
The former measures the spatial spread weighted by the number of visits to each site, while the latter records the farthest excursion from the center.

Figure~\ref{fig:loop-geometry}(a) shows that $\langle R_g\rangle$ increases from about $0.9$ lattice spacings at $n=2$ to $1.25$ at $n=6$, while the mean contour length increases by approximately a factor of 16. The curves for $d=2,3,4$ nearly coincide. The maximum excursion grows more noticeably, reaching about $2.4$--$2.6$ lattice spacings at $n=6$ [Fig.~\ref{fig:loop-geometry}(b)], with the largest values in $d=2$ and the smallest in $d=4$. Thus, both the typical spread and the farthest excursion grow much more slowly than the contour length over the sampled depths, although the outermost visited positions extend farther than suggested by $R_g$ alone.

The distribution of spatial variance among directions is described by the gyration tensor~\cite{solc1971,aronovitz1986,rudnick1986}
\begin{equation}
G_{\mu\nu}=\frac{1}{\ell}\sum_{k=0}^{\ell-1}
(r_{k,\mu}-r_{\mathrm{cm},\mu})
(r_{k,\nu}-r_{\mathrm{cm},\nu}),
\label{eq:loop_gyration_tensor}
\end{equation}
whose trace is $R_g^2$. Its eigenvalues $\lambda_\mu$ define the effective dimension~\cite{recanatesi2022,hill1973}
\begin{equation}
	D_{\mathrm{eff}} =\frac{(\operatorname{Tr}G)^2}{\operatorname{Tr}(G^2)} 
	=\frac{\left(\sum_{\mu=1}^{d}\lambda_\mu\right)^2} {\sum_{\mu=1}^{d}\lambda_\mu^2}.
\label{eq:loop_effective_dimension}
\end{equation}
This quantity approaches one when a single direction dominates and equals $d$ when the variance is shared equally among all directions. Figure~\ref{fig:loop-geometry}(c) shows that $\langle D_{\mathrm{eff}}\rangle$ remains near $1.8$ in two dimensions but increases with depth in three and four dimensions. The spatial distribution therefore becomes more balanced among the available directions without a comparable increase in radius. Here $D_{\mathrm{eff}}$ measures shape anisotropy; it is neither a fractal dimension nor a measure of spatial filling.

\begin{figure*}[t]
\centering
\includegraphics[width=\textwidth]{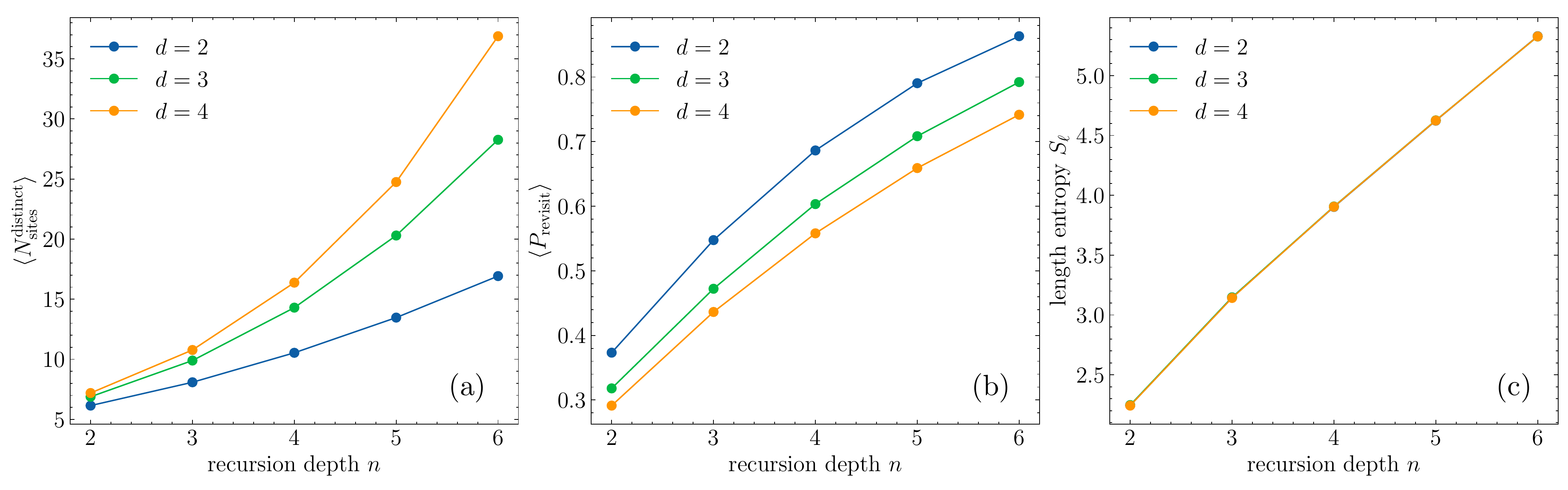}
\caption{Site revisitation and loop-length diversity as functions of recursion depth $n$ for dimensions $d=2$, $3$, and $4$. (a) Mean number of distinct lattice sites visited by a loop. (b) Mean revisitation fraction $\langle P_{\mathrm{revisit}}\rangle$, where $P_{\mathrm{revisit}}=1-N_{\mathrm{sites}}^{\mathrm{distinct}}/\ell$ for a closed path of length $\ell$, with the repeated final endpoint omitted. (c) Shannon entropy of the loop-length distribution, $S_\ell(n)=-\sum_\ell P_n(\ell)\ln P_n(\ell)$. Although the number of distinct sites grows with depth, the increasing revisitation fraction shows that many additional steps return to previously visited sites. The entropy curves nearly coincide across dimensions, consistent with the same length-generation rules in each dimension.}
\label{fig:loop-entropy}
\end{figure*}

To distinguish exploration of new sites from repeated traversal, we define the number of distinct visited sites of a closed path $\gamma$ as~\cite{rudnick2004}
\begin{equation}
N_{\mathrm{sites}}^{\mathrm{distinct}}(\gamma)
=\left|\{\mathbf r_0,\ldots,\mathbf r_{\ell-1}\}\right|.
\label{eq:loop_distinct_sites}
\end{equation}
Each site contributes only once to this count, regardless of its number of visits. The fraction of visits that return to a previously encountered site is
\begin{equation}
P_{\mathrm{revisit}}(\gamma)
=1-\frac{N_{\mathrm{sites}}^{\mathrm{distinct}}(\gamma)}{\ell}.
\label{eq:loop_revisit}
\end{equation}
The plotted mean $\langle P_{\mathrm{revisit}}\rangle$ is obtained by averaging this fraction over individual loops, rather than by taking a ratio of ensemble averages.

Figures~\ref{fig:loop-entropy}(a,b) show that deeper loops visit more distinct sites while devoting an increasing fraction of their traversals to revisitation. At $n=6$, the mean distinct-site count is roughly 17, 28, and 37 in $d=2,3,4$, respectively. The revisitation fraction is largest in two dimensions and reaches about $0.74$ even in four dimensions. Unlike the length distribution, these exploration statistics depend strongly on dimension: additional spatial directions allow more distinct sites to be visited despite similar mean lengths and radii. Repeated traversal is partly built into the message rules. Node products concatenate loops based at a common site, while the transported node message follows an edge path and then its exact reverse, necessarily revisiting the sites on that path.

The diversity of contour lengths is quantified by the Shannon entropy~\cite{shannon1948}
\begin{equation}
S_\ell(n)=-\sum_{\ell}P_n(\ell)\ln P_n(\ell).
\label{eq:loop_length_entropy}
\end{equation}
Figure~\ref{fig:loop-entropy}(c) shows approximately linear growth over the depths studied, with nearly coincident curves in different dimensions, as expected from the dimension-independent length distribution. The associated effective number of populated length sectors is~\cite{hill1973,jost2006}
\begin{equation}
N_{\mathrm{len}}(n)=\exp[S_\ell(n)].
\label{eq:loop_effective_lengths}
\end{equation}
Its approximately exponential increase characterizes the broadening of $P_n(\ell)$. Geometrically different paths of the same length belong to the same sector, so $N_{\mathrm{len}}$ does not count distinct loop shapes or independent Wilson-loop operators.

Correlations between contour length and individual-loop geometry are examined in Appendix~\ref{app:loop_correlations}. They complement the depth-dependent averages by showing the variation in spatial extent, site exploration, and shape among loops of comparable length.

Together, these results establish a separation between exponential growth of the mean contour length and much slower growth of spatial extent over the sampled depths. The measured length describes the full generated path before algebraic simplification. Backtracking cancellations and trace identities can relate paths of different lengths, so contour growth does not by itself quantify the growth of an independent operator basis. The geometric statistics characterize the structures produced by the specified recursion; their weights in a trained network or a physical observable require additional information.

\section{Implications for learning lattice gauge observables}
\label{sec:implications}

The separation between contour length and spatial extent has direct implications for the representation of lattice gauge observables. A simple example is provided by fermions hopping in a fixed gauge background,
\begin{equation}
\begin{aligned}
\widehat H[U]&=t\sum_{\langle ij\rangle}\left(c_i^\dagger U_{ij}c_j+\mathrm{H.c.}\right)\\
&=\sum_{ij}c_i^\dagger h_{ij}[U]c_j,
\end{aligned}
\label{eq:gauge_matter_hamiltonian}
\end{equation}
where $c_i$ is an $N_c$-component fermion operator, $U_{ij}\in\mathrm{SU}(N_c)$, and each undirected nearest-neighbor bond is counted once. For this quadratic Hamiltonian, the thermal energy is~\cite{goedecker1994,goedecker1999,wang2018}
\begin{equation}
E[U;\mu,T]=\operatorname{Tr}\!\left[h[U]f_\beta(h[U]-\mu)\right],
\label{eq:fermion_energy}
\end{equation}
with $f_\beta(x)=(e^{\beta x}+1)^{-1}$ and $\beta=1/(k_{\mathrm B}T)$. The trace runs over sites and internal components. At fixed mean filling, the chemical potential is determined by $\operatorname{Tr}f_\beta(h[U]-\mu)=N_f$ and can depend on the gauge configuration.

The connection to loops follows from the spectral moments of the hopping matrix,
\begin{equation}
\begin{aligned}
\operatorname{Tr}h^\ell&=t^\ell\sum_{\gamma\in\mathcal C_\ell}\operatorname{tr}_c U_\gamma,\\
U_\gamma&=U_{i_0i_1}U_{i_1i_2}\cdots U_{i_{\ell-1}i_0},
\end{aligned}
\label{eq:hopping_loop_moments}
\end{equation}
where $\mathcal C_\ell$ contains rooted, oriented closed walks of length $\ell$, and $\operatorname{tr}_c$ acts on the internal indices. These walks include backtracking, repeated site visits, and multiple windings. We denote their summed traces by 
\begin{eqnarray}
	\mathcal W_\ell[U]=\sum_{\gamma\in\mathcal C_\ell}\operatorname{tr}_c U_\gamma. 
\end{eqnarray}
The reverse orientations are included in this sum, making $\mathcal W_\ell$ real.

Explicit expansions are particularly simple at half filling on a bipartite lattice. With hopping only between the two sublattices, the spectrum of $h$ is symmetric about zero and $\mu=0$. Expanding the Fermi function at high temperature gives
\begin{equation}
\begin{aligned}
E[U;0,T]
={}&-\frac{\beta t^2}{4}\mathcal W_2[U]
+\frac{\beta^3t^4}{48}\mathcal W_4[U]\\
&-\frac{\beta^5t^6}{480}\mathcal W_6[U]+\cdots.
\end{aligned}
\label{eq:energy_high_temperature}
\end{equation}
The series converges when $\beta\|h\|<\pi$. Deep in this regime, higher hopping orders are suppressed and short loops dominate. The length-two walks are gauge independent because $U_{ij}U_{ji}=I$; on a hypercubic lattice, the first gauge-dependent contributions arise from elementary plaquettes at length four. Equation~\eqref{eq:energy_high_temperature} thus gives a direct example of an observable organized by contour length, with coefficients fixed by the physical parameters.

At zero temperature and exact half filling, the occupied negative-energy states instead give
\begin{equation}
E_0[U]=-\frac{1}{2}\operatorname{Tr}|h[U]|.
\label{eq:energy_half_filling_zero_temperature}
\end{equation}
The nonanalyticity of $|\epsilon|$ at zero prevents a Taylor expansion about that point, but a convergent polynomial representation is obtained using Chebyshev polynomials~\cite{silver1994,weisse2006,wang2018}. Choose a configuration-independent energy scale $a>\|h\|$ and write $\widetilde h=h/a$. For $N_s$ lattice sites, the Chebyshev expansion of the absolute-value function yields
\begin{equation}
E_0[U]=-\frac{aN_cN_s}{\pi}
+\frac{2a}{\pi}\sum_{m=1}^{\infty}
\frac{(-1)^m}{4m^2-1}\operatorname{Tr}T_{2m}(\widetilde h),
\label{eq:energy_zero_temperature_chebyshev}
\end{equation}
where $T_m$ is the Chebyshev polynomial of the first kind. Each term combines closed-walk sums of even lengths up to $2m$. For example,
\begin{equation}
\begin{aligned}
\operatorname{Tr}T_2(\widetilde h)
&=\frac{2t^2}{a^2}\mathcal W_2-N_cN_s,\\
\operatorname{Tr}T_4(\widetilde h)
&=\frac{8t^4}{a^4}\mathcal W_4-\frac{8t^2}{a^2}\mathcal W_2+N_cN_s.
\end{aligned}
\label{eq:chebyshev_loop_combinations}
\end{equation}
The coefficients in Eq.~\eqref{eq:energy_zero_temperature_chebyshev} decay algebraically as $m^{-2}$, reflecting the cusp in the energy function. They weight polynomial combinations of loop lengths, rather than individual length sectors; retaining this grouping is essential to the convergence of the representation. These two examples illustrate how the organization and weighting of loop contributions depend on the observable and physical regime.

The closed-walk expansions distinguish two aspects of electronic propagation: contour length specifies hopping order, whereas spatial extent describes the region explored. Repeated returns allow high-order contributions to remain within a small neighborhood. Thus, a finite locality range does not require a truncation at short contour length. The gauge products generated by message passing have the same algebraic form as these contributions, although their sampling probabilities differ from the physical weights determined by the observable and its parameters.

To make the meaning of locality precise, consider the exact site decomposition
\begin{equation}
E=\sum_i\mathcal E_i[U],
\qquad
\mathcal E_i[U]=\operatorname{tr}_c\!\left[h f_\beta(h-\mu)\right]_{ii}.
\label{eq:local_fermion_energy}
\end{equation}
This choice assigns half of each bond's energy to each endpoint. Although the associated operator involves only adjacent bonds, its expectation value depends on the electronic state and hence generally on the entire gauge configuration. A site decomposition therefore does not, by itself, establish local dependence on the gauge field.

Electronic nearsightedness supplies the additional physical principle: local expectation values become insensitive to sufficiently distant changes of the Hamiltonian at fixed chemical potential~\cite{Kohn1996,Prodan2005}. For bounded finite-range hopping at finite temperature, the analyticity of the Fermi function permits exponentially accurate local approximations~\cite{benzi2013}. Restricting the Hamiltonian to the graph-distance-$R$ neighborhood of $i$, at the same $\mu$ and $T$, gives a bound of the form
\begin{equation}
\left|\mathcal E_i-\mathcal E_i^{(R)}\right|\leq C e^{-R/\xi},
\qquad
R\geq \xi\ln(C/\varepsilon),
\label{eq:local_energy_accuracy}
\end{equation}
where the second relation is sufficient for a local error no larger than $\varepsilon<C$. The loop interpretation is direct: leaving this neighborhood and returning requires at least $2R+2$ steps, so local moments through degree $2R+1$ are unchanged by the restriction. Polynomial convergence then bounds the effect of the omitted exterior, while arbitrarily long walks remain possible inside the neighborhood. At zero temperature, a spectral gap supports exponential density-matrix decay, whereas gapless metals generally exhibit algebraic tails~\cite{ismailbeigi1999,benzi2013}. At fixed filling, configuration dependence of the chemical potential introduces an additional global dependence.

The separation between contour length and spatial extent observed in Fig.~\ref{fig:loop-geometry} is consistent with this distinction between propagation order and physical range. Long paths can describe repeated propagation through nearby links, without requiring sensitivity to distant gauge fields. Their importance nevertheless depends on the physical regime. In the controlled high-temperature expansion, higher orders are suppressed and short loops can capture most of the gauge-dependent energy. At lower temperatures, many orders may contribute even when a finite neighborhood suffices for the desired accuracy. Compactness alone does not enhance a walk's contribution, which depends on its holonomy, expansion coefficient, and cancellations with other contributions.

The measured geometry therefore characterizes possible propagation structures rather than their physical importance. Contour lengths are counted before algebraic simplification: immediate backtracking can cancel, and tracing a transported loop gives $\operatorname{tr}_c(PWP^{-1})=\operatorname{tr}_c W$. Moreover, the sampled $R_g$ does not determine the physical locality length $\xi$. The central physical implication is that spatial locality and high-order gauge-field dependence are compatible; the loop weights determine which of the available structures contribute appreciably to a given observable.

\section{Conclusion and Outlook}
\label{sec:conclusion}

We have investigated the statistical geometry of lattice paths generated by gauge-equivariant graph neural networks. By tracing the recursive composition of node and edge features, we formulated a stochastic construction of the closed walks available through message passing, independently of network training. Local feature multiplication concatenates existing loops, while gauge-covariant transport combines structures anchored at neighboring sites. These operations generate paths with repeated visits, retraced segments, and nested excursions. The associated length recursions and probability-generating functions provide an analytical description of the loop-length distribution, including the exponential growth of its mean, $\langle\ell\rangle\sim(14/5)2^n$. Under the specified sampling rules, this distribution is independent of spatial dimension, even though the geometry of the generated paths depends on dimension.

The central geometrical result is the separation between contour growth and spatial expansion. Over the depths studied, both the radius of gyration and maximum excursion increase much more slowly than the mean contour length. Deeper loops explore additional sites but devote an increasing fraction of their steps to revisitation, while the growing length-distribution entropy describes an increasingly broad range of contour lengths. The gyration-tensor analysis further shows that paths can distribute their spatial variance more evenly among the available directions without becoming substantially larger. Correlations between length and geometry reveal considerable variation among loops of comparable length, confirming that hopping order alone does not determine spatial extent or shape. Together, these results characterize a hierarchy of increasingly elaborate gauge-link products whose additional length is accumulated largely through repeated traversal of nearby regions.

For GNN design, this hierarchy distinguishes spatial reach from the complexity represented within that reach. Nearest-neighbor updates obey the receptive-field bound~\cite{gilmer2017,reiser2022}
\begin{equation}
R_{\mathrm{rec}}(n)\leq R_{\mathrm{init}}+n,
\label{eq:receptive_field_bound}
\end{equation}
where $R_{\mathrm{init}}$ accounts for the initial feature support. This maximum reach is compatible with much slower growth of the typical loop radius. The latter characterizes the distribution of visited positions rather than the full receptive field or the spatial sensitivity of a trained model. Sufficient spatial coverage remains necessary for observables involving distant links, including extended Wilson loops.

The same distinction applies to learning local observables, including action densities, local fermionic expectation values, and forces associated with individual sites or links. Whenever the dependence on distant inputs is sufficiently weak at the desired accuracy, a local predictor can be written schematically as
\begin{equation}
\mathcal O_i[U]\simeq\mathcal O_{i,\theta}[U]
=\mathcal F_\theta\!\left(\mathcal N_i^{(R)}[U]\right),
\label{eq:local_observable_ml}
\end{equation}
where $\mathcal N_i^{(R)}[U]$ denotes the supplied gauge environment and the output obeys the transformation law of the target quantity. An analogous construction applies to quantities anchored on links. The locality assumption concerns the dependence on the surrounding gauge configuration; it does not follow simply from assigning the observable to a site or link.

The complexity of this dependence varies substantially among observables. The Wilson gauge-action density depends explicitly on elementary plaquettes, whereas local fermionic quantities, such as the chiral-condensate density, involve the fermion propagator and can receive contributions from repeated propagation through the surrounding gauge field~\cite{wilson1974,rothe2012}. Forces obtained after integrating out fermions can likewise depend on many propagation processes. Even when a finite neighborhood gives an accurate approximation, the dependence within that neighborhood may therefore require high-order combinations of nearby links. The observed separation between contour length and spatial extent identifies how additional depth can build such combinations after the necessary spatial coverage has been reached. Its usefulness is consequently controlled by both the range and the complexity of the target observable.

The benefit of this additional composition depends on the observable and physical regime. In the controlled high-temperature example, short loops can already capture most of the gauge-dependent energy, so further depth may provide diminishing returns. When many hopping orders contribute appreciably, deeper layers may instead help represent repeated propagation within the relevant neighborhood. Carry-forward messages retain access to shorter structures, allowing learned combinations to select among the available orders rather than necessarily emphasizing the longest paths.

The recursion reaches mean formal hopping orders of size $L$ at depths $n=\log_2 L+O(1)$ through reuse of intermediate features, suggesting an efficient alternative to explicitly enumerating long-loop inputs. Formal contour growth alone, however, does not determine the number of independent invariants represented at finite width or establish an improvement in prediction accuracy. An important next step is to examine how training selects and combines the available structures, and whether improvements with depth arise primarily from access to more distant gauge fields or from more elaborate combinations within the existing spatial support. Relating these learned representations to the physical loop contributions would clarify how the geometrical hierarchy identified here supports the prediction of observables with different locality ranges and degrees of complexity.

\begin{acknowledgments}
This work was supported by the U.S. Department of Energy, Office of Basic Energy Sciences, under Contract No.~DE-SC0020330, which supported the development of the ML and GNN methodology. The application to lattice gauge theories and related conceptual developments were motivated in part by the EXCLAIM collaboration, supported by the U.S. Department of Energy under Contract No.~DE-SC0024644. The authors thank Yaohang Li for collaboration on related work, M.~Engelhardt and H.-W.~Lin for insightful discussions on lattice gauge theories, and S.~Liuti for discussions in connection with the EXCLAIM collaboration. The authors also acknowledge Research Computing at the University of Virginia for computational resources and technical support.
\end{acknowledgments}

\appendix

\section{Recursive generation of gauge loops}
\label{app:loop_generation}

\begin{figure*}[t]
\centering
\includegraphics[width=\textwidth]{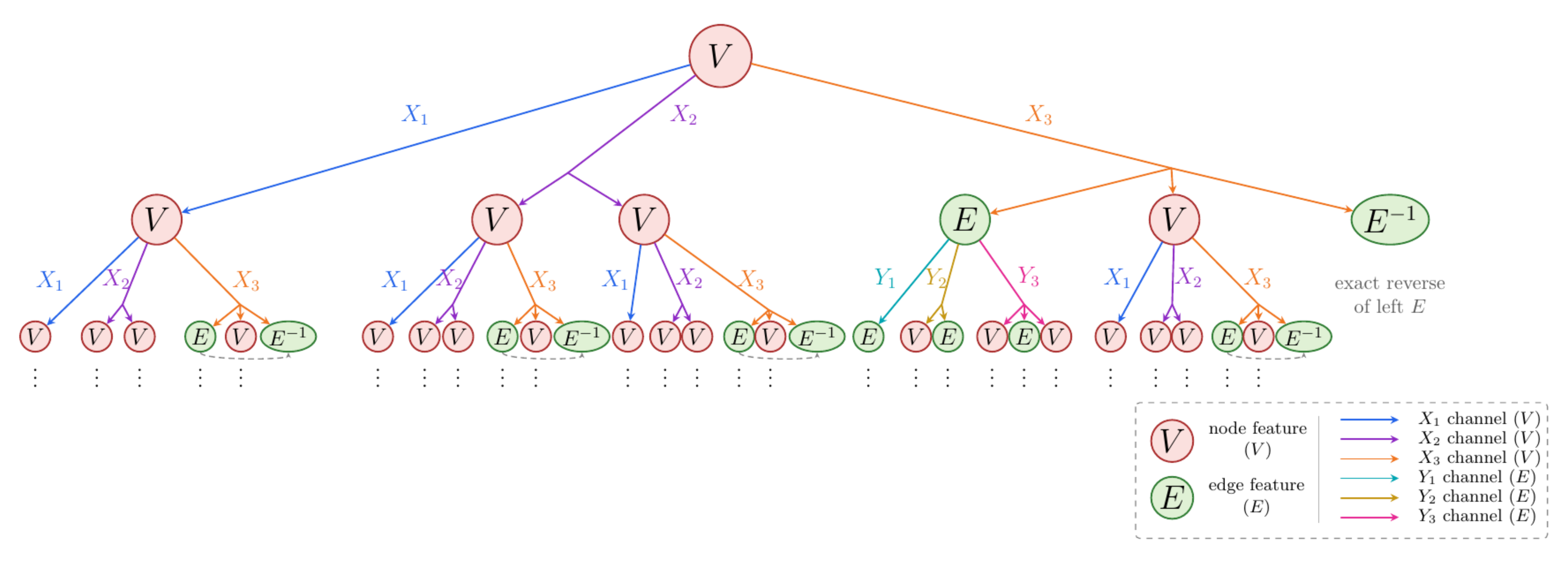}
\caption{Recursive generation of closed paths by gauge-equivariant message passing. The tree traces the possible origins of a node feature $V$ at the output layer through successive message-passing layers. The $X_1$, $X_2$, and $X_3$ branches carry forward a node feature, multiply two node features, and transport a neighboring node feature between two edge features, respectively [Eqs.~\eqref{eq:node_message_1}--\eqref{eq:node_message_3}]. The $Y_1$, $Y_2$, and $Y_3$ branches carry forward an edge feature $E$ or dress it with node features at one or both endpoints [Eqs.~\eqref{eq:edge_message_1}--\eqref{eq:edge_message_3}]. Choosing an operation at each branch, together with all factors required by that operation, specifies an ordered product of input features. The resulting node feature represents a closed lattice path, whose trace is a gauge-invariant loop observable. The dots indicate that the recursion continues to the input links and plaquettes.}
\label{fig:tree-schematic}
\end{figure*}

We generate the loop ensemble by recursively selecting individual contributions to the node and edge messages defined in Eqs.~\eqref{eq:node_message_1}--\eqref{eq:node_message_3} and \eqref{eq:edge_message_1}--\eqref{eq:edge_message_3}. The procedure follows the algebraic structure of message passing but does not require a trained network or its learned channel weights. Beginning with a node feature at recursion depth $n$, we select one message contribution, recursively construct every factor in that contribution, and continue until reaching the input links and plaquettes. Each realization thus specifies one ordered product contributing to an output node feature. Its trace is a gauge-invariant loop observable.

In this appendix, $V$ and $E$ denote individual path contributions to the feature collections $\mathbf V$ and $\mathbf E$, while $\mathbf X_r$ and $\mathbf Y_r$ label the message types with site and layer indices suppressed. For an individual edge holonomy, path reversal gives $E^\dagger=E^{-1}$ because it is a product of unitary links; this inverse relation need not hold for a general learned edge feature.

Figure~\ref{fig:tree-schematic} illustrates this construction. A node feature $V$ has three possible origins, corresponding to the $\mathbf{X}_1$, $\mathbf{X}_2$, and $\mathbf{X}_3$ messages. The $\mathbf{X}_1$ operation retains one node feature from the preceding depth, while $\mathbf{X}_2$ requires two node features at the same site. The $\mathbf{X}_3$ operation selects a neighboring site $j$ and requires three factors: an edge feature connecting $i$ to $j$, a node feature at $j$, and the transporter along the same edge in the reverse direction. An edge feature $E$ similarly branches through the $\mathbf{Y}_1$, $\mathbf{Y}_2$, and $\mathbf{Y}_3$ messages. Here $\mathbf{Y}_2$ contains two distinct products, depending on which endpoint supplies the node feature. Consequently, the three edge-message channels correspond to four possible products in the stochastic construction. Selecting one operation at each encountered feature and retaining all factors required by that operation forms a \emph{derivation subtree} of the full schematic tree. This is analogous to an AND/OR derivation tree: choosing a message operation selects one alternative, while constructing the chosen product requires all of its factors~\cite{ChangSlagle1971}. Together with the choices of sites and elementary paths, this subtree specifies one generated loop. A single branch from the root to an input feature traces only one factor's ancestry and does not specify the complete loop.

For each node object at depth $n>0$, we choose the $\mathbf{X}_1$, $\mathbf{X}_2$, or $\mathbf{X}_3$ operation with equal probability. When $\mathbf{X}_3$ is chosen, we also select one of the $2d$ nearest neighbors with equal probability on a $d$-dimensional hypercubic lattice. For each edge object, we choose uniformly among the four products represented by $\mathbf{Y}_1$, the two terms of $\mathbf{Y}_2$, and $\mathbf{Y}_3$. The factors required by the chosen product are then generated recursively at depth $n-1$. Separate factors are sampled independently, with one essential exception: the two edge factors in $\mathbf{X}_3$ describe the same transporter traversed in opposite directions. We generate that path once and use its exact reverse for the return journey.


The recursion terminates at $n=0$. A node object is initialized as an elementary plaquette. We choose two distinct coordinate directions $\mu$ and $\nu$ and signs $s_\mu,s_\nu=\pm1$; its ordered sequence of nearest-neighbor displacements is
\begin{equation}
\bigl(s_\mu\hat{\mathbf e}_\mu,\,
s_\nu\hat{\mathbf e}_\nu,\,
-s_\mu\hat{\mathbf e}_\mu,\,
-s_\nu\hat{\mathbf e}_\nu\bigr).
\label{eq:loop_initial_plaquette}
\end{equation}
An edge object at depth zero is the elementary link between its specified endpoints. Its orientation follows that of the edge object being constructed. Thus the input node and edge objects provide, respectively, closed paths and open paths connecting neighboring sites.

We store a generated path as an ordered sequence of unit displacements $\gamma=(\mathbf d_1,\ldots,\mathbf d_{L_\gamma})$, where $\mathbf d_k\in\{\pm\hat{\mathbf e}_1,\ldots,\pm\hat{\mathbf e}_d\}$. Products of features concatenate their corresponding paths in the order dictated by matrix multiplication. For the transported node message, let $P_{i\to j}$ denote the generated path from $i$ to the selected neighbor $j$. The resulting closed path has the form
\begin{equation}
\begin{aligned}
\gamma_i &= P_{i\to j}\circ\gamma_j\circ P_{i\to j}^{-1},\\
(\mathbf d_1,\ldots,\mathbf d_m)^{-1} &= (-\mathbf d_m,\ldots,-\mathbf d_1),
\end{aligned}
\label{eq:loop_inverse_path}
\end{equation}
where $\gamma_j$ is the recursively generated loop based at $j$ and $\circ$ denotes path concatenation. In particular, the return path is determined by $P_{i\to j}$ and is never sampled independently.

Every generated node path is closed. This follows inductively: an input plaquette is closed; $\mathbf{X}_1$ preserves a closed path; $\mathbf{X}_2$ concatenates two paths based at the same site; and $\mathbf{X}_3$ follows an open path to a neighboring site, traverses a closed path there, and returns along the exact inverse open path. The edge operations likewise preserve the specified endpoints of an open path. The construction permits self-intersections, repeated visits to sites, and repeated traversals of links. We retain the full ordered path, including immediately retraced links; any cancellation used to simplify a schematic drawing is not part of the generation algorithm. Recursion depth therefore controls the number of rounds of possible composition, rather than fixing the contour length or spatial extent of a loop.

The replacement rules also determine the probability distribution of contour lengths without sampling explicit lattice paths. Let $P_n(\ell)$ be the probability that a node object at depth $n$ generates a path of length $\ell$, and let $Q_n(\ell)$ be the corresponding probability for an edge object. Their probability-generating functions are~\cite{rudnick2004}
\begin{equation}
F_n(z)=\sum_\ell P_n(\ell)z^\ell,
\qquad
G_n(z)=\sum_\ell Q_n(\ell)z^\ell.
\label{eq:loop_length_pgf_def}
\end{equation}
Products of independently generated factors contribute products of generating functions. The three equally likely node replacements and four equally likely edge products give
\begin{eqnarray}
& & F_n(z) = \frac{1}{3}\bigl[F_{n-1}(z)+F_{n-1}(z)^2 +F_{n-1}(z)G_{n-1}(z^2)\bigr], \nonumber \\
& & G_n(z) = \frac{1}{4}G_{n-1}(z)[1+F_{n-1}(z)]^2.
\label{eq:loop_length_pgf}
\end{eqnarray}
In the $\mathbf{X}_3$ term, the outgoing edge and its exact reverse have the same sampled length. Their combined contribution is therefore $G_{n-1}(z^2)$, rather than $G_{n-1}(z)^2$. The elementary plaquette and link set $F_0(z)=z^4$ and $G_0(z)=z$. Iterating Eqs.~\eqref{eq:loop_length_pgf} gives $P_n(\ell)$ as the coefficient of $z^\ell$ in $F_n(z)$, providing a sampling-free benchmark for the numerical distributions in Fig.~\ref{fig:loop-length-hist}(a). The mean length discussed in the main text is the first moment $F_n'(1)$; the highest nonzero power of $F_n$ gives the maximum permitted length, which differs from the maximum observed in a finite sample. Because the elementary path lengths and replacement probabilities do not depend on $d$, these equations predict the same length distribution in every dimension considered here, although the spatial geometry of the generated paths does depend on $d$.

\section{Correlations between loop length and geometry}
\label{app:loop_correlations}

\begin{figure*}[t]
\centering
\includegraphics[width=\textwidth]{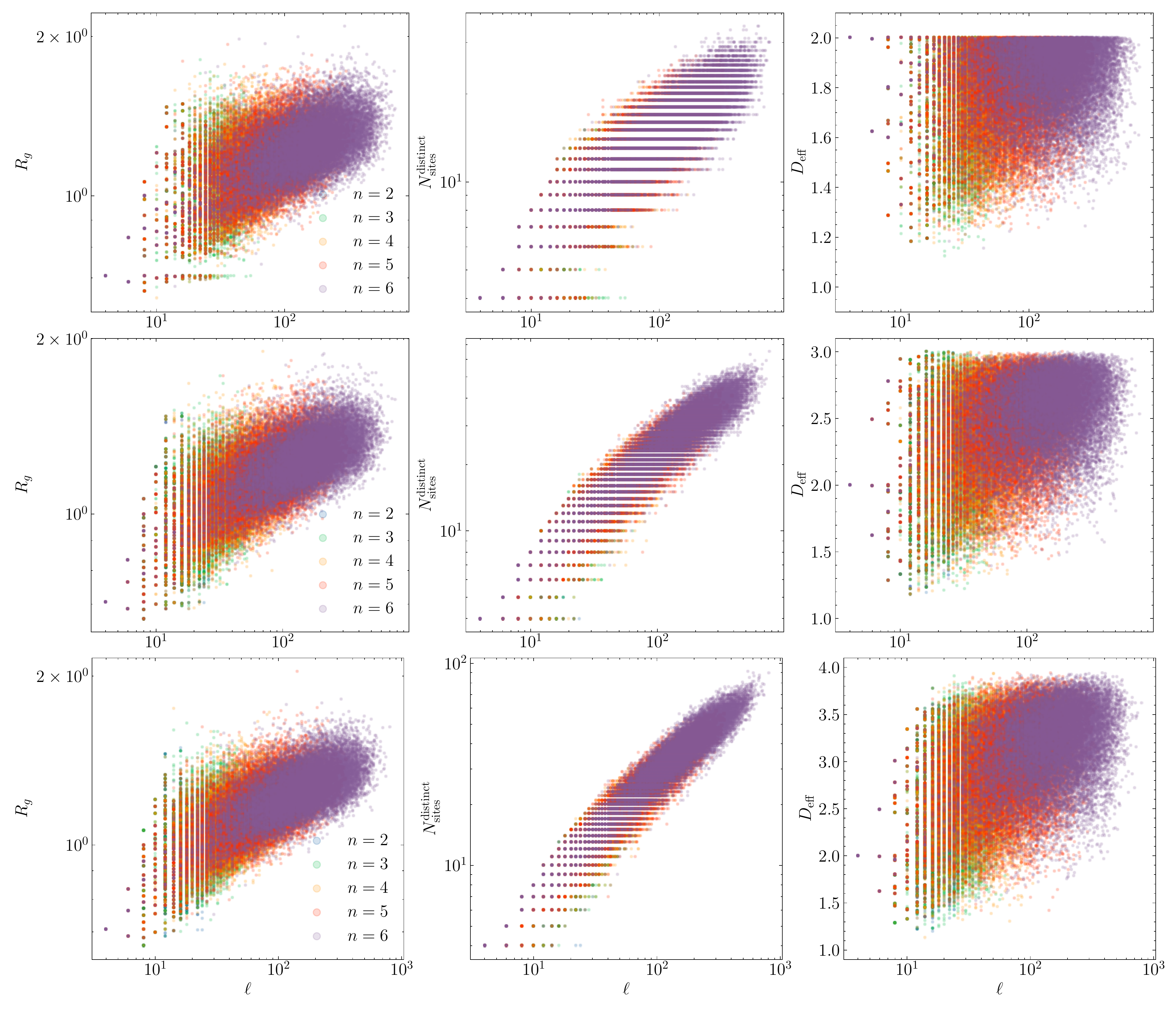}
\caption{Correlations between loop length and spatial geometry. Columns show the radius of gyration $R_g$ (left), number of distinct visited sites $N_{\mathrm{sites}}^{\mathrm{distinct}}$ (middle), and effective dimension $D_{\mathrm{eff}}$ (right) as functions of the full contour length $\ell$. Rows correspond to spatial dimensions $d=2$, $3$, and $4$, from top to bottom. Each point represents one generated loop, with colors indicating recursion depths $n=2,\ldots,6$. Longer loops generally visit more distinct sites, but their spatial extent increases only modestly, with substantial variation in geometry at a given length. The distinct-site count grows more slowly than $\ell$, indicating repeated traversal of previously visited regions. Larger $D_{\mathrm{eff}}$ corresponds to a more isotropic distribution of the visited positions.}
\label{fig:loop_geometry_vs_length}
\end{figure*}

The depth-dependent averages in Sec.~\ref{sec:loop_geometry} establish that contour length grows much faster than spatial extent. To examine how these quantities are related for individual loops, Fig.~\ref{fig:loop_geometry_vs_length} plots the radius of gyration $R_g$, number of distinct visited sites $N_{\mathrm{sites}}^{\mathrm{distinct}}$, and effective dimension $D_{\mathrm{eff}}$ against the full contour length $\ell$. Each point represents one generated loop, with colors distinguishing recursion depths $n=2,\ldots,6$ and rows corresponding to $d=2,3,4$. The definitions are those of Eqs.~\eqref{eq:loop_spatial_size}, \eqref{eq:loop_distinct_sites}, and \eqref{eq:loop_effective_dimension}; all traversals, including repeated and retraced segments, contribute to $\ell$.

The left column shows that longer loops generally have larger $R_g$, but the increase is modest across the displayed range of contour lengths. Many paths containing hundreds of traversals still have radii of gyration of only a few lattice spacings or less. At a fixed length, the spread in $R_g$ also remains appreciable, demonstrating that the number of traversals does not uniquely determine the spatial extent. The distributions look similar across dimensions, consistent with the weak dimensional dependence of $\langle R_g\rangle$ in Fig.~\ref{fig:loop-geometry}(a). Since $R_g$ weights positions by their number of visits, it characterizes where the traversal is concentrated rather than the largest region reached by any segment of the path~\cite{rubinstein2003,rudnick2004}. The scatter supports slow growth of typical spatial extent with length, but does not establish a particular asymptotic scaling law.

The middle column shows a clearer correlation between $\ell$ and the number of distinct visited sites. Longer loops generally encounter more sites, while the distinct-site count increases more slowly than the total number of traversals over the displayed range. Their additional length thus combines exploration of new sites with repeated visits to sites already encountered. The dimensional dependence is more pronounced than for $R_g$: at comparable contour lengths, loops in higher dimensions typically visit more distinct sites despite similar characteristic radii. Additional spatial directions allow a larger set of sites to be explored within a region of comparable typical extent. These correlations provide an individual-loop counterpart to the dimensional dependence of the site count and revisitation fraction in Fig.~\ref{fig:loop-entropy}(a,b).

The right column describes the shape of the spatial distribution through $D_{\mathrm{eff}}$. In two dimensions, the longer-loop population is concentrated toward $D_{\mathrm{eff}}=2$, although substantial variation remains. In three and four dimensions, longer loops also tend toward larger effective dimensions, indicating that the spatial variance is shared more evenly among the available directions. This change accompanies only a modest increase in $R_g$, so a loop can become more isotropic without becoming substantially larger. As in the main text, $D_{\mathrm{eff}}$ measures the balance of the gyration-tensor eigenvalues~\cite{recanatesi2022}; it does not determine a fractal dimension or imply that all sites within the spatial envelope are visited.

The populations associated with different recursion depths broadly overlap in all three columns. Increasing depth extends the range of accessible lengths, while loops from different depths can have comparable lengths and geometries. This overlap suggests that contour length accounts for part of the depth dependence of the geometric averages, but it does not establish that the conditional geometry at fixed $\ell$ is independent of $n$. The scatter therefore distinguishes the broadening of the length distribution from the variation of spatial structure among loops of the same length.

Taken together, the panels show how long paths accumulate traversals within a slowly expanding spatial region while continuing to explore additional sites and spatial directions. This behavior is consistent with the recursive construction: node products concatenate loops at a common base point, and transported node messages follow an outgoing path, traverse a neighboring loop, and return along the exact inverse path. Repetition and nested excursions can therefore increase contour length substantially without proportional spatial expansion. The correlations reinforce the main distinction between the order of a generated gauge-link product and its spatial range, while showing that loops of a given order retain considerable geometric diversity.

\bibliography{ref.bib}

\end{document}